\documentclass{svproc}
\usepackage{url}
\usepackage{amsmath}
\usepackage{amssymb}
\usepackage{graphicx}
\newcommand*{\rom}[1]{\expandafter\@slowromancap\romannumeral #1@}

\begin{document}
\mainmatter              
\title{Few-body insights of multiquark exotic hadrons}
\titlerunning{Few-body insights of multiquark exotic hadrons}  
%
\author{Javier Vijande\inst{1} \and Jean-Marc Richard \inst{2} \and Alfredo Valcarce\inst{3}}
\authorrunning{Javier Vijande et al.} 
%
\tocauthor{Javier Vijande, Jean-Marc Richard, Alfredo Valcarce}
\institute{Unidad Mixta de Investigaci\'on en Radiof\'\i sica e Instrumentaci\'on 
Nuclear en Medicina (IRIMED), Instituto de Investigaci\'on Sanitaria La Fe 
(IIS-La Fe)-Universitat de Valencia (UV) and IFIC (UV-CSIC), Valencia, Spain
\and
Universit\'e de Lyon, Institut de Physique Nucl\'eaire de Lyon,
IN2P3-CNRS--UCBL,
4 rue Enrico Fermi, 69622  Villeurbanne, France
\and
Departamento de F{\'\i}sica Fundamental and IUFFyM,
Universidad de Salamanca, 37008 Salamanca, Spain
}

\maketitle              

\begin{abstract}
In this contribution we discuss the adequate treatment of the $4-$ and $5-$body dynamics within a constituent quark framework. 
We stress that the variational and Born-Oppenheimer approximations give energies rather close to the exact ones, while 
the diquark approximation might be rather misleading. Hall-Post inequalities provide very useful lower bounds that 
exclude possible stable states for some mass ratios and color wave functions.  

\end{abstract}
\section{Introduction}

Recent contributions on multiquarks are stimulated by the discovery of a double-charm baryon~\cite{Aai17}, 
which is interesting by itself and also triggers speculations about exotic double-charm mesons 
$QQ\bar q\bar q$. For years, the sector of flavor-exotic tetraquarks has been somewhat forgotten, 
and even omitted from some reviews on exotic hadrons,
as much attention was paid to  hidden-flavor states $Q\bar Q q\bar q$. 
However the flavor-exotic multiquarks have been investigated already some decades 
ago~\cite{Ade82} and has motivated an abundant literature (see Ref.~\cite{Ric18} and 
references therein) that has been unfortunately ignored in some recent papers. 

In this contribution, we stress that a careful treatment of the few-body problem is 
required before drawing any conclusion about the existence of stable states in a 
particular model. Not surprisingly, the main difficulties are encountered when a 
multiquark state is found near its lowest dissociation threshold. The question of 
whether or not there is a bound state requires a lot of care.
We consider that it is important to clarify the somewhat contradictory results 
in the literature. In particular, some authors who use similar ingredients obtain 
either stability or instability for the all-heavy configuration $QQ\bar Q\bar Q$, 
and in our opinion, this is due to an erroneous handling of the four-body problem.  

\section{Diquark approximation}\label{se:diq}
A few decades ago, the main concern in baryon spectroscopy was the problem of missing 
resonances predicted by the quark model and not observed experimentally. Many states 
of the symmetric quark model disappear if baryons are constructed out of a frozen diquark 
and a quark. However, the missing resonances are not very much coupled to the typical investigation 
channels $\pi N$ or $\gamma N$, which privilege states with one pair of quarks shared with the 
target nucleon $N$. In recent photoproduction experiments with improved statistics, some of the 
missing states have been identified, which cannot be accommodated as made of a ground-state diquark 
and a third quark~\cite{Kle17}.  

The diquark model is nevertheless regularly revisited, to accommodate firmly established exotics such 
as the $X(3872)$, or even candidates awaiting confirmation. Unfortunately, some unwanted multiquarks are 
also predicted in this approach, though this is not always explicitly stated or even realized. The issue 
of unwanted multiquarks within the diquark model was raised many years ago by Fredriksson and Jandel~\cite{Fre81}, and 
is sometimes  rediscovered, without any reference to the 1982 paper. The paradox is perhaps that the diquark 
model, that produces fewer baryon states, produces too many multiquarks!

There are  many variants of the so-called diquark model. An extreme point of view is that diquarks are 
almost-elementary objects, with their specific interaction with quarks and between them. A whole baryon 
phenomenology can be built starting from well-defined assumptions about the diquark constituent masses 
and the potential linking a quark to a  diquark. Then, a diquark-diquark interaction has to be introduced as a new ingredient for the multiquark sector.  

Another extreme is to estimate the energy and wave function of, say $(a_1 a_2 a_3)$ with masses $m_i$ and interaction $v_{ij}(r)$, 
first solving for $(a_1 a_2)$ with $v_{12}$ alone with energy $\eta_{12}$, and then estimating the bound state of a 
point-like system $(a_1a_2)$ of mass $m_1+m_2$ located at $\vec R_{12}$ interacting with $a_3$ through the potential 
$v_{13}(\vec r_3-\vec R_{12})+v_{23}(\vec r_3-\vec R_{12})$, resulting in binding energy $\eta_{12,3}$. Thus, 
the whole energy is given by $\eta_{12}+\eta_{12,3}$. 

This strategy is of course fully justified for the deuterium atom considered as a $pne^-$ system, as the inter-nuclear 
motion is not significantly modified by the electron. On the other hand, this approach ruins some subtle collective binding, 
for instance, that of Borromean states~\cite{Fre06}.
Also, one cannot see either how H$^-(pe^- e^-)$ could 
become bound in this approach, or  the hydrogen molecule be described as 
a ''diproton'' linked to a ``dielectron''! In some other cases, like in the case of constituent 
quark models, the method just overestimates the binding, since the \textsl{ad-hoc} clustering lowers significantly the energy. 

For simplicity, we consider for doubly-heavy tetraquarks  the case of a frozen $\bar 33$ color wave function. 
Color mixing has to be introduced to have the proper threshold in the model, and it has been seen in explicit 
calculations that the mixing with meson-meson configurations is crucial for states at the edge of stability. 
Nevertheless the comparison of various approximations is instructive for the toy model,
\begin{equation}
\label{eq:H4}
 H_{33}=\frac{\vec p_1^2+\vec p_2^2}{2\,M}+\frac{\vec p_3^2+\vec p_4^2}{2\,m}+\frac{v(r_{12})+v(r_{34})}{2}+\frac{v(r_{13})+v(r_{14})+v(r_{23})+v(r_{24})}{4}.
\end{equation}

\begin{figure}[ht!]
 \centering
 \includegraphics[width=.5\textwidth]{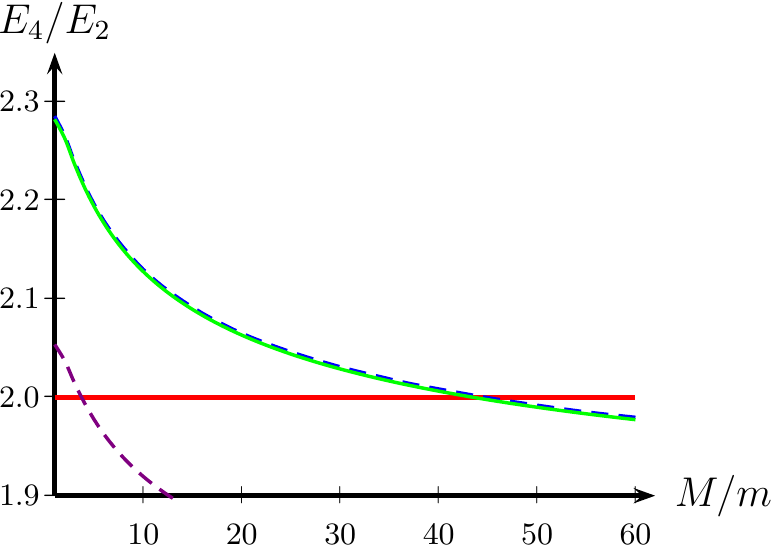}
\caption{Comparison of the variational upper bound (green curve) and Hall-Post lower bound (dotted blue curve), hardly distinguishable 
from the variational estimate at this scale, for the tetraquark Hamiltonian~\eqref{eq:H4} with a linear interaction. Also shown is the 
naive diquark-antidiquark approximation (dashed violet curve). Adapted from \cite{Ric18}.}
 \label{fig:HP1}
\end{figure}

In Fig.~\ref{fig:HP1}, we compare the exact solution of~\eqref{eq:H4} with the 
approximation consisting of first computing the $QQ$ diquark with $r_{12}/2$ alone 
and $qq$ with $r_{34}$ alone, and then $(QQ)(\bar q\bar q)$ as a meson with a potential $r_{12,34}$ and constituent masses $2\,M$ and $2\,m$. 

\section{Relating mesons, baryons and tetraquarks}\label{se:EQ}
In a recent paper, Eichten and Quigg~\cite{Eic17} use heavy-quark symmetry to 
relate meson, baryon and tetraquark energies. In a simplified version without spin effects, it reads
\begin{equation}
 \label{eq:EQ1}
 QQ\bar q\bar q=QQq + Qqq-Q\bar q~,
\end{equation}
where the configuration stands for the ground-state energy. For fixed $m$ and $M\to \infty$, the identity 
is exact. For finite $M$, there is some departure. 
If one treats the tetraquark $QQ\bar q\bar q$ and the doubly-heavy baryon $QQq$ in the  Born-Oppenheimer 
approximation, one can compare the two effective potentials as a function of the $QQ$ separation $x$, the 
baryon one being shifted by $ Qqq-Q\bar q$ which is independent of $x$. Without recoil correction, the two 
potentials are identical at $x=0$. For finite $M$, there is slight difference, as the single $q$ recoils 
against either $M$ or $2\,M$, and similarly $qq$ recoils against one or two heavy quarks.

\begin{figure}[ht]
 \centering
\includegraphics[width=.5\textwidth]{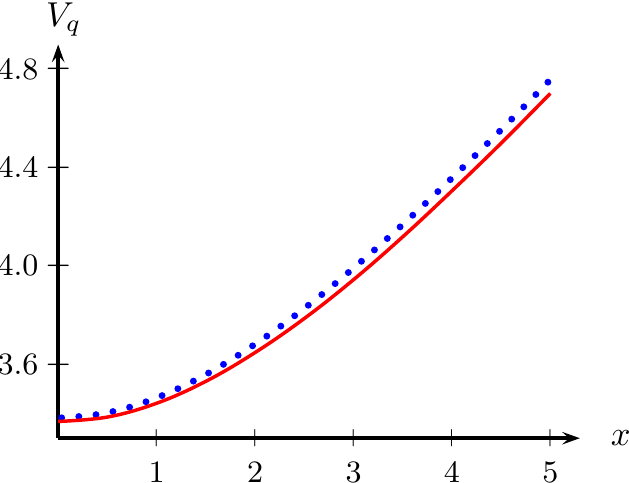}
 \caption{Comparison of the light quark energies for $QQ\bar q\bar q$ (solid red line) and 
$QQq$ (dotted blue line) as a function of the $QQ$ separation~$x$. The second curve is shifted by the 
difference of energies $Qqq-Q\bar q$. The units are such $m=1$, $M=5$ and $v_{ij}=r_{ij}$. Adapted from \cite{Ric18}.}
 \label{fig:BO34}
\end{figure}

The comparison is shown in Fig.~\ref{fig:BO34}. Clearly the two effective potentials are very 
similar, and thus give almost identical energies, up to an additive constant that corresponds to the last two terms in~\eqref{eq:EQ1}.

\section{Hall-Post inequalities}\label{se:HP}
The Hall-Post inequalities have been derived in the 50s to relate the binding energies 
of light nuclei with different number of nucleons~\cite{Hal67}. They have been 
rediscovered in the course of studies on the stability of matter~\cite{Fis66}, or to 
link meson and baryon masses in the quark model~\cite{Ade82,Nus99}.  

For their application to tetraquarks we consider the toy Hamiltonian~\eqref{eq:H4}. It can be rewritten as,
\begin{eqnarray}\label{eq:HP12}
&&\left(\sum\vec p_i\right).(A(\vec p_1+\vec p_2)+B(\vec p_3+\vec p_4))
+\frac{\tilde h_{12}(x_{12})+\tilde h_{34}(x_{34})}{2}+\sum{\strut}'{\,}\frac{\tilde{\tilde h}_{ij}(x,a,b)}{4}~,\nonumber\\
&&\tilde{\tilde h}_{13}(x,a,b)=\frac1x\,\genfrac{(}{)}{}{}{\vec p_1-\vec p_3+ a\,\vec p_2+b\,\vec p_4}{2}^2+v_{ij}~,
 \end{eqnarray}
where the masses $x_{12}$, $x_{34}$ and  $x$ are readily calculated from the parameters $A$, $B$ and $a$, and $b$.
This results into 
\begin{equation}\label{eq:HP13}
 E_4(M,m)\ge \max_{A,B,a,b}\left[E_2(x_{12})+ E_2(x_{34})+E_2(x)\right]~.
\end{equation}
Hence a rigorous lower bound is obtained from simple algebraic manipulations and the knowledge of the 2-body 
energy as a function of the reduced mass. For a linear interaction the results for $E_4/E_2(1)$ as a function 
of $M/m$ are shown in Fig.~\ref{fig:HP1}. The sum $1/M+1/m$ is kept equal to 2 to fix the threshold energy at $2\,E_2(1)$. 

\section{Color mixing and spin-dependent corrections}\label{se:col-mix}
Any model  with a pairwise potential, due to color-octet exchange, induces mixing between $\bar33$ and $6\bar6$ 
states in the $QQ-\bar q\bar q$ basis. Perhaps the true dynamics inhibits the call for higher color representations 
such as sextet, octet, etc., for the subsystems of a multiquarks, but for the time being, let us adopt the color-additive 
model. If one starts from a $\bar33$ state with $QQ$ in a spin triplet, and, for instance $\bar q\bar q=\bar u\bar d$ with 
spin and isospin $S=I=0$, then its orbital wave function is mainly made of an $s$-wave in all coordinates. It can mix with 
a color $6\bar 6$ with orbital excitations in the $\vec x$ and $\vec y$ linking $QQ$ and $\bar q\bar q$, respectively. 

To illustrate the role of color-mixing for the AL1 potential we use the potential AL1 by Semay 
and Silvestre-Brac~\cite{Sem94}. Its central part is a Coulomb-plus-linear term, while its 
spin-spin part is a regularized Breit-Fermi interaction with a smearing parameter that depends on the reduced mass. 

\begin{figure}[ht]
 \centering
\includegraphics[width=.45\textwidth]{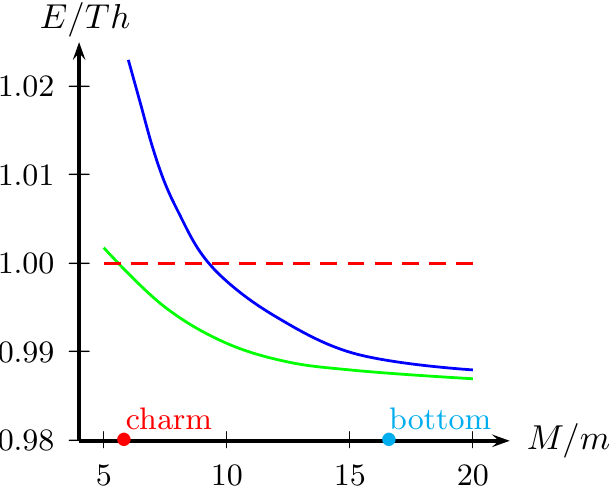}
\includegraphics[width=.45\textwidth]{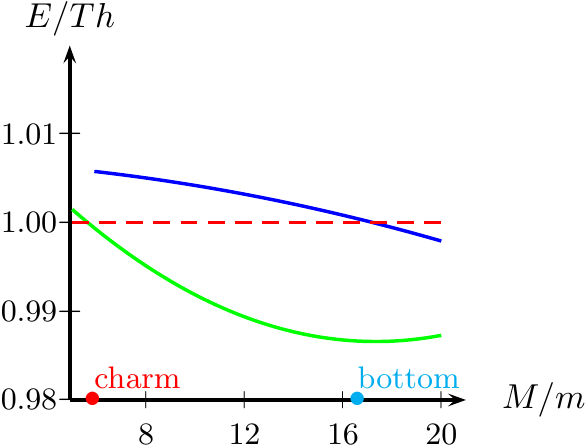}
\caption{(left) Effect of color-mixing on  the binding of $QQ\bar u\bar d$, within the AL1 model. The tetraquark energy is 
calculated with only the color $\bar33$ configurations (blue curve) and with the $6\bar6$ components (green curve). (right) 
Effect of the spin-spin interaction of the binding of $QQ\bar u\bar d$, within the AL1 model. The tetraquark energy is 
calculated with (green line) and without (blue line) the chromomagnetic term. Adapted from \cite{Ric18}.}
 \label{fig:AL1-col-mix}
\end{figure}

The energy, normalized to the lowest threshold, as a function of $M/m$ without and with color-mixing is shown in Fig.~\ref{fig:AL1-col-mix}(left). 
The ground state of the $QQ\bar u\bar d$ with $J^P=1^+$, which is a candidate for stability, has its main component 
with color $\bar33$, and spin $\{1,0\}$ in the $QQ-\bar u\bar d$ basis. The main admixture consists of $6\bar 6$ with 
spin $\{1,0\}$ and an antisymmetric orbital wavefunction, and of $6\bar 6$ with spin $\{0,1\}$ with a symmetric orbital wavefunction.

It has been acknowledged in the literature that a pure additive interaction such as~\eqref{eq:H4} will not bind $cc\bar q\bar q$, 
on the sole basis that this tetraquark configuration benefits from the strong $cc$ chromoelectric attraction that is absent in 
the $Q\bar q+Q\bar q$ threshold. In the case where $qq=ud$ in a spin and isospin singlet, however, there is in addition a favorable chromomagnetic interaction 
in the tetraquark, while the threshold experiences only heavy-light spin-spin interaction, whose strength is suppressed by a factor $m/M$. 

To study the spin-dependent corrections, we make use once again of the AL1 potential. The results are shown in 
Fig.~\ref{fig:AL1-col-mix} (right) for $QQ\bar u\bar d$, as a function of the mass ratio $M/m$.

The system $bb\bar u\bar d$ is barely bound without the spin-spin term, though the mass ratio $m_b/(m_u\simeq m_d)$ is very large. 
It acquires its binding energy of the order of 150\,MeV~\cite{Ric18} when the spin-spin is restored. 
The system $cc\bar u\bar d$ is clearly unbound when the spin-spin interaction is switched off. This is shown 
here for the AL1 model, but this is true for any realistic interaction, including an early model by Bhaduri 
et al.~\cite{Bha81}. The case of $cc\bar u\bar d$ is actually remarkable. 
Semay and Silvestre-Brac, who used their AL1 potential,  missed the binding because their method of systematic 
expansion on the eigenstates of an harmonic oscillator is not very efficient to account for the short-range 
correlations, and it was abandoned in the latest quark model calculations. Janc and Rosina were the first to 
obtain binding with such potentials, and their calculation was checked by Barnea et al. (see~\cite{Ric18} for references).

The role played by color mixing and spin-dependent corrections has also been addressed for hidden-charm pentaquarks 
$(\bar c c qqq)$~\cite{Ric18b}. The main difficulties faced when attempting to solve the 5-body problem including both 
color-mixing and spin-dependent corrections is the larger numbers of color vectors and the increased complexity of the 
radial equation. There are three independent color states for the pentaquarks: i) $(\bar c c)$ singlet coupled to $(qqq)$ 
singlet, ii) $(\bar c c)$ octet coupled  to the first $(qqq)$ octet, in which the quarks 3 and 4 are in a $\bar 3$ state, 
and iii) $(\bar c  c)$ octet associated to the second $(qqq)$ octet, in which  the quarks 3 and 4 form a sextet. Besides 
this, the larger number of coordinate ensembles and its complexity makes even more important than in the 4-quark case to 
use a reliable numerical technique to solve the radial part.  Few states were found to be bound. $(J,I)=(1/2,3/2)$ and $(3/2,3/2)$ are found below 
their lowest $S$-and $D$-wave thresholds: $\Delta \eta_c$ and $D\Sigma_c$.  In fact they are substantially lower, so that they remain 
stable or metastable even if one accounts for the width of the $\Delta$ and considers that the 
actual lowest threshold is $N \pi\eta_c$.
$(J,I)=(5/2,1/2)$ is above its lowest $D$-wave threshold while it is below
the lowest $S$-wave threshold. In this case, as highlighted long ago in Ref.~\cite{Hog78},
the contribution of color vectors different from the singlet-singlet combination would prevent 
by the centrifugal barrier the tunneling of the quarks to combine in a colorless object,
enhancing in this way the stability of this state.

\begin{figure}[t]
\begin{center}
\includegraphics[width=.55\textwidth]{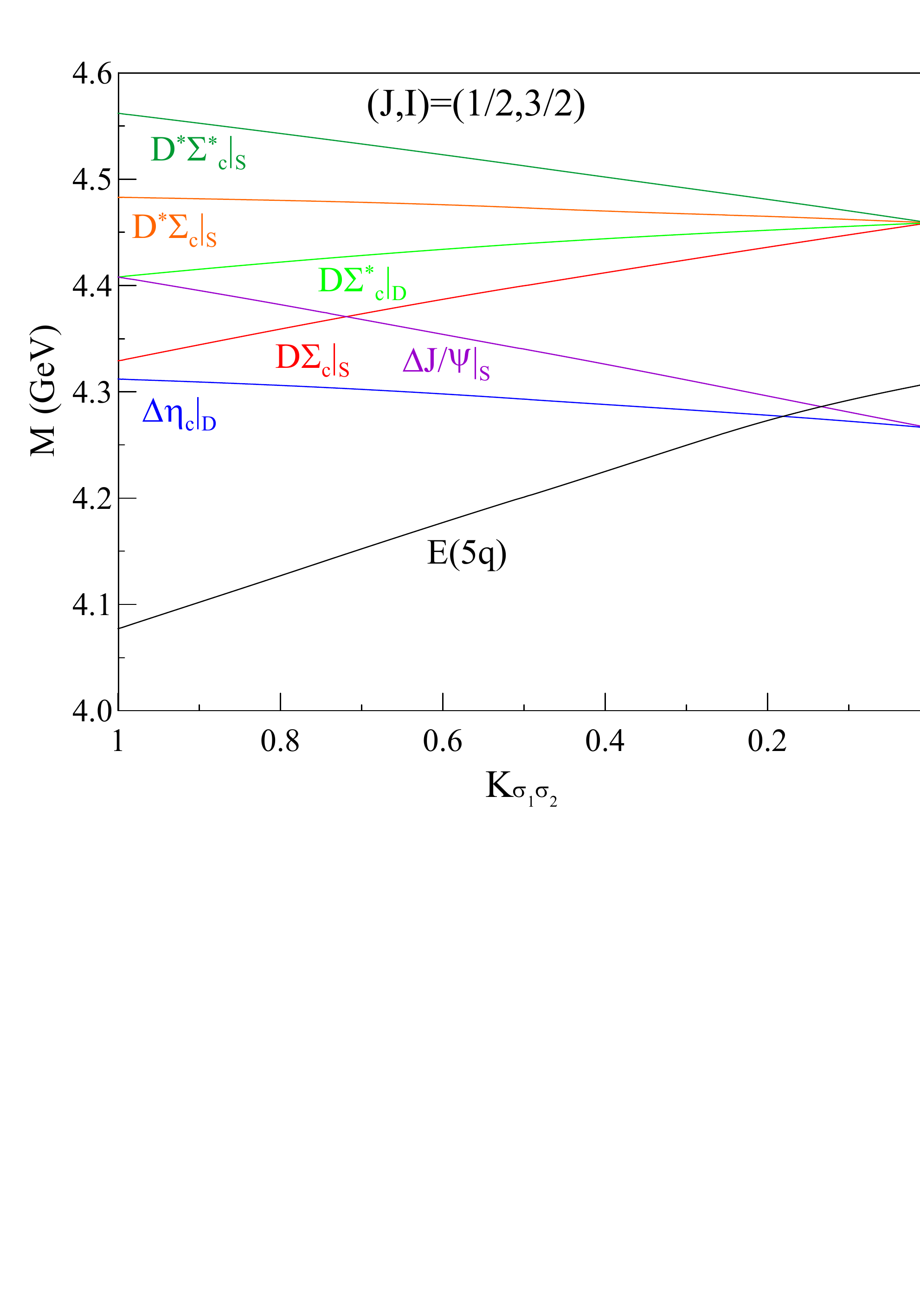}
\vspace*{-4cm}
\caption{Mass of the $(\bar c c qqq)$ $(J,I)=(1/2,3/2)$ state and its thresholds as a function
of the strength of the chromomagnetic interaction, decreased by a multiplicative factor $K_{\sigma_1\sigma_2}$. Adapted from \cite{Ric18b}.}
\label{fig:spin-spin}
\end{center}
\end{figure}

The role of the spin-spin interaction vs. the spin-independent one is shown in Fig.~\ref{fig:spin-spin} 
by modifying the strength of the chromomagnetic interaction using a multiplicative factor.  It is seen that 
the binding starts already with a small fraction 
of the spin-spin interaction. This means that there is a favorable interplay of chromoelectric 
and chromomagnetic effects, although the binding disappears in the pure chromoelectric limit.

\section{Conclusions}\label{se:concl}
Let us summarize. The few-body problem is rather delicate, especially for systems at the edge of stability.
In the case of four-quark states, the analogy with atomic physics is a good guidance to indicate the most 
favorable configurations in the limit of dominant chromoelectric interaction. However, unlike the positronium 
molecule, the all-heavy configuration $QQ\bar Q\bar Q$ is not stable if one adopts a standard quark model and solves the four-body problem correctly. 

The mixing of the $\bar 3 3$ and $6\bar 6$ color configurations is important, especially for states very near 
threshold. This mixing occurs by  both  the spin-independent and the spin-dependent parts of the potential. 

Approximations are welcome, especially if they shed some light on the four-body dynamics. The diquark-antidiquark 
approximation is not supported by a rigorous solution of the 4-body problem, but benefits of a stroke of luck, as 
the erroneous extra attraction introduced in the color $\bar 33$ channel is somewhat compensated by the neglect of 
the coupling to the color $6\bar 6$ channel. The equality relating $QQ\bar q\bar q$, $QQq$, $Qqq$ and $Q\bar q$ works 
surprisingly well as long as one is restricted to color $\bar 33$, but does not account for the attraction provided by color mixing. 
On the other hand, for asymmetric configurations $(QQ\bar q\bar q)$, the Born-Oppenheimer method provides a very good 
approximation, and an interesting insight into the dynamics. 

In short, $cc\bar u\bar d$ with $J^P=1^+$ is at the edge of binding within current quark models. For this state, 
all contributions to the binding should be added, in particular the mixing of states with different internal spin 
and color structure, and in addition, the four-body problem should be solved with extreme accuracy.
In comparison, achieving the binding of $bb\bar u\bar d$ looks easier. Still, with a typical quark 
model, the stability of the ground state below the threshold cannot be reached if spin-effects and 
color mixing are  both neglected. The crucial role of spin effects explains why one does not expect too many states besides $1^+$~\cite{Vij09}.

These effects become even more acute in the five-body case, where color-mixing and a proper 
balance between chromoelectric and chromomagnetic terms are basic to get binding. 

\section*{Acknowledgments}
This work has been partially funded by Ministerio de Econom\'\i a, Industria y Competitividad
and EU FEDER under Contract No. FPA2016-77177.

\end{document}